\documentclass{appolb}
\usepackage{epsfig}
\usepackage{amsmath}
\usepackage{axodraw}
\usepackage{graphics}

\begin{document}
\title{MULTIPLE INTERACTIONS IN LOW x DIS\\ AND HIGH ENERGY pp
COLLISIONS\footnote{Talk presented at XXXIII Intl.\ Symp.\ on
Multiplarticle Dynamics, Krak\'ow, Poland, September 2003.}}
\author{G\"osta Gustafson \\ \address{Dept. of Theoretical Physics,
Lund University\\ S\"olvegatan 14A, S-22362 Lund, Sweden\\ E-mail:
gosta@thep.lu.se} } \maketitle

\vspace*{-7cm}
\begin{flushright}
  LU-TP 03-53\\
  hep-ph/0312310\\
  November 2003
\end{flushright}
\vspace*{4.5cm}

\begin{abstract}
\noindent The LDC formalism for DIS can describe HERA structure functions, and 
also agrees well with CTEQ and MRST gluon distributions. It is also
suitable for hadronic collisions, and provides a strong connection between
$ep$ and $pp$ reactions. Some preliminary results are presented.
\end{abstract}

\section{Introduction} 
In hadron-hadron collisions \emph{collinear factorization} works well for 
calculations of high-$p_\perp$ jets. However, in this formalism the minijet 
cross section 
diverges with $\sigma_{jet} \sim 1/p_\perp^4$, which implies that also 
the total $E_\perp$ diverges. This implies the need for a soft cutoff, and 
in \textsc{Pythia}~\cite{PYTHIA} fits to experimental data give a cutoff 
$p_{\perp 0} \sim 2$ GeV. 
The cutoff is also growing with energy, which makes it difficult to 
extrapolate safely to the high energies at LHC.

In the $k_\perp$-\emph{factorization} formalism the off shell matrix element 
does not blow up when the exchanged transverse 
momentum $k_\perp\!\rightarrow\!0$. 
Assume that two incoming partons, with momenta $k_1$ and $k_2$, scatter 
producing the outgoing partons (jets) $q_1$ and $q_2$.
When the momentum exchange $k_\perp^2$ is smaller than the incoming 
virtualities $k_{\perp 1}^2$ and $k_{\perp 2}^2$, the cross section does not 
diverge,
and the total $E_\perp$ stays finite.

When $\sigma_{\mathrm{jet}} > \sigma_{\mathrm{tot}}$ it implies that there are 
often several 
hard subcollisions in a single event. Therefore correlations become important, 
and the observed ``pedestal effect'' implies that the hard subcollisions are 
not independent, indicating an impact parameter dependence such that central 
collisions have many minijets, while peripheral collisions have fewer 
minijets~\cite{zijl}.
In this talk I want to discuss connections between $ep$ and $pp$ collisions, 
and how experimental DIS data can improve predictions for high energy $pp$ 
reactions. The results presented are obtained together with Leif L\"onnblad 
and Gabriela Miu.

\section{DIS at small $x$}
\begin{figure}
\begin{center}
\scalebox{0.7}{\mbox{
\large
\begin{picture}(250,165)(0,0)
%\SetScale{0.8}
\SetOffset(50,0)
%\Gluon(100,25)(120,45){3}{4}
\Gluon(120,30)(135,61){3}{4} % \Text(130,60)[r]{ { {$k_{n-2}$} } }
\Gluon(135,61)(135,103){3}{5} \Text(137,82)[r]{ { {$k_{\perp}$} } }
\Gluon(117,133)(135,103){3}{4} %\Text(131,104)[r]{ { {$k_{n}$} } }
\Gluon(146,82)(165,82){2}{3} \Text(164,82)[l]{ { {$q_{\perp}$} } }
\Gluon(135,61)(170,54){3}{4} %\Text(158,63)[l]{ { {$q_{n-1}$} } }
\Gluon(135,103)(170,110){3}{4} %\Text(159,101)[l]{ { {$q_{n}$} } }
%\Gluon(117,118)(142,130){3}{4} \Text(140,130)[l]{ { {$q_{n+1}$} } }
%\ZigZag(90,130)(117,118){3}{5} \Text(95,133)[r]{ { {$Q^2$} } }
\Text(130,0)[]{ { {(b)} } }

\SetOffset(-190,0)
\ArrowLine(195,15)(220,15) \Text(200,20)[]{ { {$p$} } }
\Line(220,13)(250,13)
\Line(220,17)(250,17)
\Line(220,15)(240,30) \Line(240,30)(265,23)
\Gluon(240,30)(257,50){3}{4} \Gluon(257,50)(280,40){3}{4}
\Gluon(257,50)(270,70){3}{4} 
\Gluon(270,70)(295,65){3}{4} \Text(295,65)[l]{ { {$q_{\perp i}$} } }
\Gluon(270,70)(270,95){3}{4} \Text(272,82)[r]{ { {$k_{\perp i}$} } }
\Gluon(257,115)(270,95){3}{4} 
\Gluon(270,95)(295,100){3}{4} \Text(295,100)[l]{ { {$q_{\perp i+1}$} } }
\Gluon(240,135)(257,115){3}{4} \Gluon(257,115)(280,125){3}{4}
\Line(220,150)(240,135) \Line(240,135)(265,140)
\Line(220,150)(255,150)
\Photon(195,150)(220,150){3}{4} \Text(200,142)[]{ { {$\gamma^*$} } }
\BCirc(220,15){4}
\Text(245,0)[]{ { {(a)} } }
\end{picture}}}
\end{center}
\vspace*{-5mm}
\caption{(a) A large transverse momentum, 
$k_{\perp i}$, in the middle of the chain 
corresponds to a hard subcollision between a proton and a resolved photon.
(b) In the BFKL and LDC formalisms a soft emitted gluon is 
treated as final state radiation, while in the CCFM model 
some of these gluons are treated 
as initial state radiation.}
\label{figure:chain}
\end{figure}
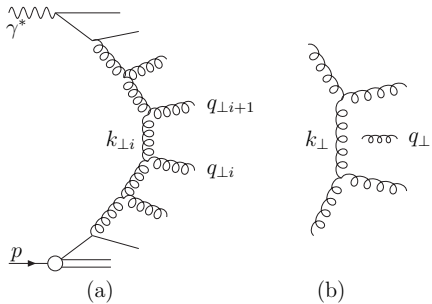

In the small $x$ (BFKL) region non-$k_\perp$-ordered parton chains are 
important. We let $k_{i}$ denote the virtual links and $q_{i}$ the (quasireal) 
emitted partons in the chain, as indicated in fig.~\ref{figure:chain}a.
Then, if the link with transverse momentum $k_{\perp i}$ is a local 
maximum, it
corresponds to a hard subcollision, where $q_{\perp i} \approx 
k_{\perp i} \approx q_{\perp i+1}$. 
A single local 
maximum corresponds to a resolved photon interaction, and with several local 
maxima there are several correlated hard subcollisions.

The BFKL equation describes only \emph{inclusive} cross sections. A 
link in the ladder corresponds to a Reggeized gluon, where soft emissions are 
compensated by virtual corrections (\emph{cf} fig.~\ref{figure:chain}b). 
Therefore such soft emissions do 
\emph{not} contribute to parton distribution functions. They \emph{do} 
contribute, 
however, to the final state properties, where they have to be added 
associated with 
appropriate Sudakov form factors.

The CCFM model interpolates between BFKL and DGLAP. Here soft emissions 
are included in the initial state radiation, with an extra suppression from 
non-eikonal form factors.
The \emph{Linked Dipole Chain} (LDC) model~\cite{LDC} is a reformulation and 
generalization of CCFM. The separation between initial and final
state radiation is here more similar to the BFKL formalism. Thus the parton 
with momentum $q$ in fig.~\ref{figure:chain}b corresponds to final state 
radiation if $q_\perp < k_\perp$, and should thus not be included in the 
initial state ladder. An important consequence is that in LDC the ISR chain is 
fully symmetric between the photon end and the proton end of the ladder.
To leading order in $\ln 1/x$ the structure function is given by
\begin{equation}
F \sim \sum_n \prod_i^n \int\int \frac{3 \alpha_s}{\pi} \frac{d z_i}{z_i}
\frac{d q_{\perp i}^2}{q_{\perp i}^2} 
\theta(q_{\perp i}-\min(k_{\perp i},k_{\perp i-1}))
\delta(x - \prod z_i)
\label{F}
\end{equation}
As $\mathbf{q}_{\perp i} \equiv \mathbf{k}_{\perp i} - \mathbf{k}_{\perp i-1}$,
the $\theta$-function implies that 
$q_{\perp i}^2 \approx \max(k_{\perp i}^2, k_{\perp i-1}^2)$. 
Expressed in the virtual link momenta 
the product of the factors $d q_{\perp i}^2/q_{\perp i}^2$
then give corresponding factors $d k_{\perp i}^2/k_{\perp i}^2$, except for 
a local maximum or minimum 
$k_{\perp i}$. For a local maximum we get a factor
$d k_{\perp i}^2/k_{\perp i}^4$, corresponding to a hard subcollision, while 
for a local minimum we just get 
$d k_{\perp i}^2$, with no singular factor. The LDC model is implemented in 
a MC event generator LDCMC~\cite{LDCMC}, which well reproduces
data on $F_2$, and also agrees well with CTEQ and MRST parametrizations for the
gluon distribution function~\cite{pdf-fit}.

\section{Hadron collisions}

The symmetry between the two ends of the parton chain implies that the LDC 
formalism also is applicable to hadron-hadron collisions. The fit to DIS data
determines the cross section for a \emph{chain} in $pp$ collisions (which 
possibly may contain more than one hard subcollision).
There is here a potential problem from the fact that with a running $\alpha_s$ 
a soft cutoff, $Q_0$, is needed. Good fits to DIS data are possible with 
different cuts, if the input distribution $f_0(x,Q_0^2)$ is adjusted 
accordingly. If the cutoff $Q_0$ is increased the number of hard 
chains (with transverse momenta $q_\perp > Q_0$) decreases, which could cause
uncertain results. However, at the same time the number of soft chains (for 
which all $q_\perp < Q_0$) increases, and as shown in ref.~ \cite{glm}, it
turns out
that the total number of chains is independent of the 
cutoff. Thus the total chain cross section $\sigma_{\mathrm{chain}} = 
\sigma_{\mathrm{hard\, chain}} + \sigma_{\mathrm{soft\, chain}}$ is fixed by
DIS data, and this implies evidently a very strong connection between DIS and 
$pp$ scattering~\cite{glm}.

There are two sources for \emph{multiple interactions}: It is possible to have 
two hard scatterings in the same chain, and there may be more than one chain 
in a single event. As described above, the LDC model can predict the 
correlations between hard scatterings within one chain, and also the average 
number of chains.
If the different chains were uncorrelated, the number of chains in one event 
would be given by a Poissonian distribution. However, events with high 
$p_\perp$ jets also have a higher background activity, the so called pedestal 
effect. This indicates that the parton subcollisions are correlated; central 
collisions have more and peripheral collisions fewer hard subcollisions.
In \textsc{Pythia} comparisons with data favour a $b$-dependence described by 
a double Gaussian distribution. The result for the number of 
subcollisions turns 
out to be very close to a geometric distribution.

\begin{figure}
\epsfig{figure=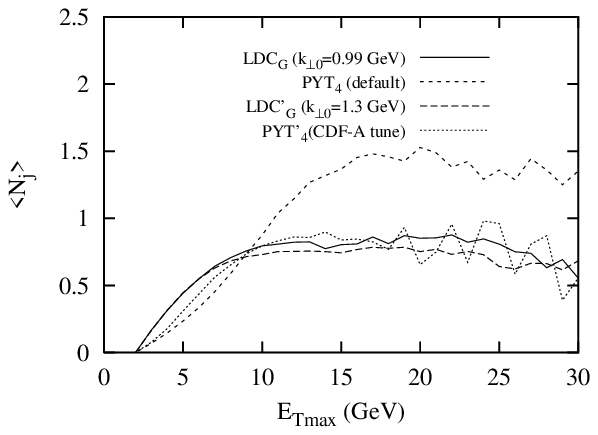,width=6.4cm}
\epsfig{figure=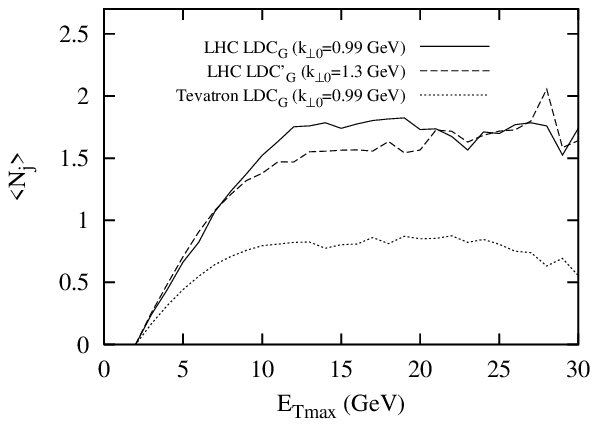,width=6.4cm}
\caption{The average number of minijets in the ``minimum azimuth 
  region'' for $|\eta|<2.5$ \emph{vs}. $E_\perp$ for the 
  hardest jet. (a) For $\sqrt s =1.8$ TeV. (b) For 14 TeV.}
\label{fig:minijets}

\end{figure}

Some preliminary results from the LDC model are shown in 
fig.~\ref{fig:minijets}. Here a geometric distribution is assumed for 
the number of chains in one event, with the tail of the distribution
reduced to satisfy
energy conservation. Fig. \ref{fig:minijets}a shows the number of minijets 
in the ``minimum azimuth region'' $60^\circ < \phi < 120^\circ$ 
at $\sqrt s =1.8$ TeV. The two
LDC curves are obtained for soft cut-off values .99 and 1.3 GeV, showing the
insensitivity to this cut-off. The two \textsc{Pythia} curves correspond to 
default parameter values, and parameters tuned to CDF data \cite{CDF}. We note 
that the LDC result agrees very well with the tuned \textsc{Pythia} result.
Fig. \ref{fig:minijets}b shows corresponding results for LHC. Also
here the two curves correspond to different cut-off values, and for comparison 
the result for 1.8 TeV is also indicated. We see that the activity increases 
by a little more than a factor of 2 between the two energies.

\vspace{5 mm}
In conclusion we have shown that there is a strong connection between DIS and 
high energy
$pp$ collisions, and preliminary results for the number of minijets and the 
pedestal effect are presented.

The symmetry of our formalism implies that the chains join at one end at 
the same rate as they multiply at the other. The chain cross section grows 
like $s^\lambda$, and therefore the average chain multiplicity satisfies 
$<\!\!n_{\mathrm{chain}}\!\!>\, \propto s^\lambda/\sigma_{\mathrm{tot}}$. 
Thus our results also may
have implications for unitarization, saturation and diffraction. Work in 
these areas is in progress.

\end{document}